\documentclass[11pt]{article}

\usepackage[preprint]{acl}
\usepackage{tabularx}
\usepackage{times}
\usepackage{latexsym}
\usepackage[T1]{fontenc}
\usepackage[utf8]{inputenc}
\usepackage{microtype}
\usepackage{inconsolata}
\usepackage{graphicx}
\usepackage{booktabs}
\usepackage{multirow}
\usepackage{amsmath}
\usepackage{amssymb}
\usepackage{xcolor}
\usepackage{colortbl}
\usepackage{pifont}
\usepackage{tikz}
\usetikzlibrary{arrows.meta,positioning}

\newcommand{\bench}{\textsc{EmoShift}}   
\newcommand{\cmark}{\ding{51}}
\newcommand{\xmark}{\ding{55}}
\newcommand{\AB}{\textsc{a$\to$b}}

\renewcommand{\AA}{\textsc{a$\to$a}}

\title{Too Emotional to Be Fake?
Intra-Utterance Emotion Shifts \\ Expose Failure Modes
in Speech Deepfake Detectors}

\author{
  Jingyi Chen \\
  Reality Defender \\
  \texttt{jingyi@realitydefender.ai}
  \And
  Surya Koppisetti \\
  Reality Defender \\
  \texttt{surya@realitydefender.ai}
}

\begin{document}
\maketitle


\begin{abstract}
Expressive speech synthesis can now change
emotion within a single utterance, yet its
impact on deepfake detection remains unclear.
We introduce \bench{}, a benchmark comparing
emotion-stable (\AA{}) and emotion-changing
(\AB{}) speech to examine these effects while
accounting for static emotion identity and
acoustic confounds.
It comprises 7,240 bona fide utterances from
four source datasets and 9,720 synthetic
utterances generated using four expressive
TTS systems.
Across seven detectors, we find that emotion
changes can make synthetic speech appear more
genuine, increasing spoof miss rates by up to
25.3 percentage points on OMG-Emotion at fixed
thresholds.
Conversely, emotion-changing bona fide speech
increases false-alarm rates by up to 24.4
percentage points on IEMOCAP.
Further analyses reveal detector-specific
sensitivities to emotion trajectories, not
fully explained by static emotion identity
or splicing artifacts.
These findings underscore the importance of
emotion dynamics in deepfake detection.
\end{abstract}


\section{Introduction}
\label{sec:intro}

Emotion is rarely static in natural speech.
Within a single utterance, a speaker may shift
from calm to angry, sadness to excitement, or
fluctuate between more subtle emotional states.
Advances in expressive text-to-speech
(TTS) have made such dynamics possible in
synthetic speech
\citep{chen25b_interspeech,zhou2025indextts2,liang2026tedtts}.
As synthetic voices become more expressive,
we ask:
\emph{how do speech deepfake detectors respond
to emotion changes within an utterance?}

Existing benchmarks provide limited insight
into this question. Standard speech deepfake
benchmarks
\citep{wang2020asvspoof,wang2024asvspoof5,muller2022itw}
focus primarily on the authenticity label.
Emotion-focused benchmarks such as EmoFake
\citep{zhao2024emofake} and AffectDF
\citep{mahapatra2026affectdf} evaluate detection
across emotion labels, but treat emotion
primarily as an utterance-level property.
However, differences in detection performance
across emotion categories cannot reveal whether
detectors are sensitive to changes in emotion
within an utterance, beyond the individual
emotions being expressed.
Addressing this question requires comparing
emotion-stable and emotion-changing speech
while considering other factors that may
affect detector responses.

To investigate this question, we introduce
\bench{}, a benchmark contrasting emotion-stable
(\AA{}) and emotion-changing (\AB{}) speech
in both bona fide and synthetic audio.
Evaluating both authenticity classes allows
us to examine whether emotion changes make
synthetic speech harder to detect, bona fide
speech more prone to false alarms, or both.
The benchmark contains 16,960 utterances,
including 7,240 bona fide utterances sourced
from four publicly available datasets and
9,720 synthetic utterances generated using
four expressive TTS systems.
Our design combines controlled same-speaker,
same-text comparisons in ESD~\citep{zhou2021esd}
with naturally occurring emotional variation
in OMG-Emotion~\citep{barros2018omg},
IEMOCAP~\citep{busso2008iemocap}, and
MSP-Podcast~\citep{lotfian2019msp}.

We characterize emotion using both categorical
labels (e.g., happy or sad) and continuous
valence, arousal, and dominance (VAD)
annotations \citep{mehrabian1974basic},
representing emotional pleasantness,
activation, and perceived control.
While categorical labels identify discrete
emotional states, VAD captures continuous
differences in emotion, including variation
within the same category.
Together, these annotations characterize
both categorical emotion changes and
continuous differences in emotional states,
including their magnitude and direction.
We compare detector scores separately for
bona fide and synthetic speech, examine
detection errors at fixed thresholds, and
analyze potential confounding factors.

We evaluate seven speech deepfake detectors
on \bench{} and identify distinct detection
failures. On OMG-Emotion, all seven detectors
assign lower spoof scores to emotion-changing
synthetic speech, increasing spoof miss rates
by 9.7--25.3 percentage points at thresholds
calibrated on emotion-stable speech
(\S\ref{sec:q1}; Figure~\ref{fig:shift},
Appendix Table~\ref{tab:full}).
In contrast, emotion-changing bona fide speech
increases false-alarm rates by up to 24.4
percentage points on IEMOCAP
(\S\ref{sec:q2}).
These effects depend on the detector, dataset,
and emotion trajectory: XLS-R-based detectors
tend to assign lower spoof scores to
emotion-changing synthetic speech on ESD
and OMG-Emotion, whereas AASIST and RawNet2
respond strongly to increasing arousal in
bona fide speech
(\S\ref{sec:q3}; Figure~\ref{fig:emotion}).
Additional analyses suggest that static emotion
identity and splicing artifacts alone do not
fully explain these patterns
(\S\ref{sec:q4}).

Our contributions are threefold:
\begin{itemize}
    \setlength\itemsep{0pt}
    \item We introduce \bench{}, a benchmark
    combining controlled and naturalistic
    comparisons of emotion-stable (\AA{})
    and emotion-changing (\AB{}) bona fide
    and synthetic speech across four source
    datasets.
    \item We systematically evaluate the effects
    of emotion changes through class-specific
    detector score shifts, fixed-threshold
    detection errors, and analyses of potential
    confounding factors.
    \item We reveal detector-specific responses
    to emotion changes, showing that they can
    increase spoof miss rates or false-alarm
    rates depending on speech conditions and
    emotion trajectories.
\end{itemize}
The \bench{} benchmark and
demo audio samples are available on our project website
(\url{https://emoshift.github.io/emoshift/}).

\section{Related Work}

\paragraph{Benchmarks and generalization.}
ASVspoof \citep{wang2020asvspoof,wang2024asvspoof5}
established EER-based evaluation on studio-controlled read
speech; In-the-Wild \citep{muller2022itw}
demonstrated substantial performance degradation
on real-world recordings.
Recent work highlights the challenges of
generalizing deepfake detection beyond benchmark
conditions \citep{berisha2025generalize}
and the limitations of EER at deployment
thresholds \citep{firc2026lastmile}.
We therefore also evaluate detection errors
at fixed thresholds.

\paragraph{Emotion and expressive attacks.}
EmoFake \citep{zhao2024emofake} uses emotional
voice conversion to modify the emotion of ESD
utterances, while AffectDF
\citep{mahapatra2026affectdf} evaluates emotional
attacks across TTS, VC, and audio-LLM generators.
Both benchmarks primarily examine utterance-level
emotion. Recent advances in fine-grained emotion
control allow synthetic speech to change emotion
within an utterance
\citep{liang2026tedtts,
liu2026emotrattssmoothintrautteranceemotion,chen-etal-2026-audio},
raising new questions about how deepfake detectors
respond to such changes.

\paragraph{Temporal structure within an utterance.}
PartialSpoof \citep{zhang2023partialspoof}
studies temporal changes in authenticity by
embedding synthetic segments into bona fide
utterances. In contrast, our work examines
temporal changes in emotion while authenticity
remains constant throughout each utterance.
Prior research has characterized emotion as
a dynamic process
\citep{russell1980circumplex,kuppens2017dynamics},
but its implications for speech deepfake
detection remain underexplored.



\section{The \bench{} Benchmark}
\label{sec:bench}

\bench{} contains 16,960 utterances, including
7,240 bona fide utterances constructed from four
source datasets and 9,720 synthetic utterances
generated using four expressive TTS systems
(Table~\ref{tab:benchmark}).
The benchmark is organized around comparisons
between emotion-stable (\AA{}) and
emotion-changing (\AB{}) speech in both
authenticity classes, combining controlled
and naturalistic evaluation settings.


\begin{table*}[t]
\centering
\setlength{\tabcolsep}{3.5pt}
\renewcommand{\arraystretch}{1.08}
\resizebox{\textwidth}{!}{%
\begin{tabular}{@{}l l l c l l r r c c@{}}
\toprule
 & Speech & Emotion & Intra-utt. &
 Comparison & &
 \multicolumn{2}{c}{Utterances} &
 \multicolumn{2}{c}{Duration (s), \AA{} / \AB{}} \\
\cmidrule(lr){7-8}
\cmidrule(l){9-10}
Dataset & style & labels & transition &
 design & Generators &
 Bona fide & Spoof & Bona fide & Spoof \\
\midrule

\multicolumn{10}{@{}l}{\emph{Related benchmarks}} \\

ASVspoof 2019 LA
& read & -- & \xmark & --
& TTS, VC & 12{,}483 & 108{,}978 & --& --\\

In-the-Wild
& spontaneous & -- & \xmark & --
& in-the-wild & 19{,}963 & 11{,}816 & --& --\\

PartialSpoof
& read & -- & \xmark & --
& TTS, VC & 12{,}483 & 108{,}978 & --& --\\

EmoFake
& acted & categorical & \xmark & --
& EVC & 35{,}000 & 72{,}800 & -- & -- \\

AffectDF
& acted + spont. & categorical & \xmark & --
& TTS, VC, EVC, LALM
& 19{,}108 & 266{,}689 & --& --\\

\bench{} (ours)
& acted + spont. & categorical + VAD & \cmark
& controlled + natural
& TTS & 7{,}240 & 9{,}720
& 10.16 / 10.33 & 10.68 / 10.87 \\

\midrule

\multicolumn{10}{@{}l}{\emph{\bench{} composition}} \\

ESD
& acted, read & categorical & \cmark
& same speaker \& text
& I, T, H & 800 & 2{,}400
& 5.69 / 5.70 & 5.13 / 5.18 \\

OMG-Emotion
& in-the-wild & categorical & \cmark
& separate utterances
& I, T, H & 440 & 1{,}320
& 11.49 / 11.38 & 6.89 / 7.08 \\

IEMOCAP
& dyadic & categorical + VAD & \cmark
& separate utterances
& E & 1{,}600 & 1{,}600
& 8.85 / 9.21 & 10.42 / 10.91 \\

MSP-Podcast
& podcast & categorical + VAD & \cmark
& separate utterances
& E & 4{,}400 & 4{,}400
& 11.32 / 11.48 & 14.95 / 15.09 \\

\bottomrule
\end{tabular}}

\caption{
Comparison with related benchmarks (top) and
\bench{} composition (bottom).
ESD uses same-speaker, same-text \AA{}/\AB{}
pairs; other datasets use separate utterances.
Generators: I = IndexTTS2, T = TED-TTS,
H = HybridEmo, E = EmoTra-TTS.
Durations are mean utterance lengths (s)
for \AA{}/\AB{} within each class.
}

\label{tab:benchmark}
\end{table*}


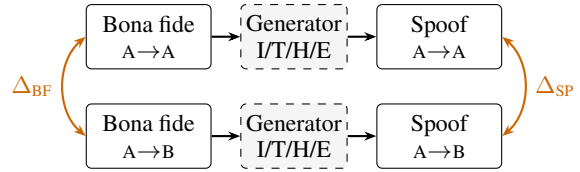
\begin{figure}[t]
\centering
\begin{tikzpicture}[
  font=\footnotesize,
  box/.style={
    draw,
    rounded corners=2pt,
    minimum width=1.65cm,
    minimum height=0.85cm,
    align=center,
    inner sep=2pt
  },
  gen/.style={
    draw,
    dashed,
    rounded corners=2pt,
    minimum width=1.3cm,
    minimum height=0.85cm,
    align=center,
    fill=black!4,
    inner sep=2pt
  },
  arr/.style={
    -{Stealth[length=1.6mm]},
    thick
  },
  cmp/.style={
    {Stealth[length=1.6mm]}-{Stealth[length=1.6mm]},
    thick,
    color=orange!80!black
  }
]

\node[box] (bfA) {Bona fide\\\AA{}};
\node[box, below=0.45cm of bfA]
  (bfB) {Bona fide\\\AB{}};

\node[gen, right=0.38cm of bfA]
  (gA) {Generator\\I/T/H/E};

\node[gen, right=0.38cm of bfB]
  (gB) {Generator\\I/T/H/E};

\node[box, right=0.38cm of gA]
  (spA) {Spoof\\\AA{}};

\node[box, right=0.38cm of gB]
  (spB) {Spoof\\\AB{}};

\draw[arr] (bfA) -- (gA);
\draw[arr] (gA) -- (spA);

\draw[arr] (bfB) -- (gB);
\draw[arr] (gB) -- (spB);

\draw[cmp] (bfA.west)
  to[bend right=50]
  node[left] {$\Delta_{\text{BF}}$}
  (bfB.west);

\draw[cmp] (spA.east)
  to[bend left=50]
  node[right] {$\Delta_{\text{SP}}$}
  (spB.east);

\end{tikzpicture}

\caption{
Benchmark construction.
Each spoof follows the text and emotion targets
of its bona fide source.
$\Delta_{\text{BF}}$ and $\Delta_{\text{SP}}$
measure the (\AB{})--(\AA{}) score difference
within each class.
}
\label{fig:design}
\end{figure}

\subsection{Benchmark Construction}
\label{sec:design}
\label{sec:corpora}
\label{sec:generators}

Each utterance contains two
speech regions with emotional states
$(e_1,e_2)$. Emotion-stable (\AA{}) speech
maintains the same emotion across both regions,
whereas emotion-changing (\AB{}) speech changes
from emotion $A$ to $B$ ($A\neq B$).

Constructing these comparisons poses two
challenges. First, speakers rarely produce
the same utterance with different emotional
trajectories in naturally occurring speech,
making matched comparisons difficult to obtain.
Second, existing public speech datasets do not
readily provide annotated within-utterance
emotion changes for systematic evaluation.
We therefore combine controlled and naturalistic
settings with different levels of experimental
control.

Under each setting, we construct both bona fide
and synthetic speech in the \AA{} and \AB{}
conditions (Figure~\ref{fig:design}).
Synthetic utterances are generated using
the transcripts and region-level emotion
targets of their bona fide sources.

\paragraph{Controlled: ESD.}
ESD \citep{zhou2021esd} contains parallel recordings
of the same sentences in five emotions: neutral,
angry, happy, sad, and surprise.
For each of 10 English speakers, we construct
\AA{}/\AB{} pairs by concatenating two sentences.
Both versions share an identical first recording
expressing emotion $A$.
Their second regions contain the same sentence
spoken by the same speaker, but express emotion
$A$ in \AA{} and emotion $B$ in \AB{}.
We obtain 400 pairs, balanced across all 20
ordered emotion transitions (20 pairs each).

\paragraph{In-the-wild: OMG-Emotion.}
OMG-Emotion \citep{barros2018omg} contains emotional
monologues from online videos.
We select 220 \AA{} excerpts with two consecutive
utterances sharing the same emotion label,
and 220 \AB{} excerpts where the label changes.
All excerpts are continuous and unspliced.
Unlike ESD, \AA{} and \AB{} are different
source utterances with different texts.

\paragraph{Naturalistic: IEMOCAP and MSP-Podcast.}
These conversational datasets provide continuous
valence, arousal, and dominance (VAD) annotations
\citep{busso2008iemocap,lotfian2019msp}.
Each example combines two speech regions from
the same speaker. We use VAD annotations to
measure differences in emotional states between
the two regions, including differences within
the same categorical emotion.
Specifically, we compute the Euclidean distance
$d_{\mathrm{VAD}}$ between their VAD vectors,
with each dimension normalized to $[0,1]$.

For IEMOCAP and MSP-Podcast, \AA{} requires the
same categorical emotion and a small
$d_{\mathrm{VAD}}$, while \AB{} requires different
emotion categories and a large $d_{\mathrm{VAD}}$.
The selected ranges are 0.00--0.13 (\AA{})
versus 0.84--1.09 (\AB{}) for IEMOCAP, and
0.00--0.14 versus 0.55--1.40 for MSP-Podcast.
The final subsets contain 800/800 \AA{}/\AB{}
examples from IEMOCAP and 2,200/2,200 from
MSP-Podcast.
Full selection criteria are provided in
Appendix~\ref{sec:appendix-data}.

\paragraph{Synthetic speech generation.}
We generate spoofs using four state-of-the-art expressive TTS systems
that differ in how they control emotion.
For each bona fide source, we use its
transcript and region-level emotion
annotations as generation targets.
IndexTTS2 \citep{zhou2025indextts2} synthesizes
each region separately, conditioning on its
corresponding emotion label and a reference voice
from the source speaker, before concatenating
the two outputs.
TED-TTS \citep{liang2026tedtts} uses single-pass,
segment-aware emotion control, assigning the
source emotion labels to their corresponding
text segments while conditioning on the source
speaker's voice.
HybridEmo
\citep{zhou2026sequentialtrajectoriessimultaneousblending}
generates the full utterance from a natural-language
instruction specifying the sequence of emotions
in the source, together with its transcript and
speaker reference.
We apply all three generators to both ESD and
OMG-Emotion, generating one synthetic utterance
per bona fide source with each generator.

For the naturalistic corpora, we use EmoTra-TTS
\citep{liu2026emotrattssmoothintrautteranceemotion}, built on CosyVoice~2
\citep{du2024cosyvoice2}, to generate speech
conditioned on source transcripts and continuous
valence--arousal--dominance (VAD) trajectories.
VAD annotations from both source regions
define the target emotional trajectory,
enabling single-pass generation of emotion-stable
or emotion-changing speech.
All audio is resampled to 16\,kHz before scoring.

\subsection{Addressing Potential Confounding Factors}
\label{sec:confounds}

\paragraph{Static emotion identity.}
Differences between \AA{} and \AB{} may
reflect the emotions expressed in each region,
rather than emotional change alone.
In ESD, we balance all 20 ordered transitions
to maintain a balanced distribution of
starting and ending emotions.
For OMG-Emotion, we compare each \AB{}
synthetic sample against the average score
of emotion-stable synthetic samples expressing
its starting and ending emotions, within
the same generator.
This analysis examines whether static
emotion identity alone accounts for the
observed score differences
(\S\ref{sec:q4}).

\paragraph{Linguistic content.}
We address potential linguistic-content
confounds through complementary controlled
and naturalistic comparisons.
In ESD, \AA{} and \AB{} share the same
speaker and text, with identical first-region
audio, providing a controlled comparison.
In OMG-Emotion, IEMOCAP, and MSP-Podcast,
\AA{} and \AB{} are constructed from
different source utterances, reflecting
naturally occurring emotional variation.
We analyze detector score shifts separately
for bona fide and synthetic speech and,
for naturalistic bona fide speech, additionally
adjust for speaker, duration, and regional
VAD annotations
(Appendix~\ref{sec:appendix-nat}).
These complementary analyses allow us to
examine emotion-related detector responses
across settings with different levels of
experimental control.

\paragraph{Audio splicing.}
Concatenation can introduce acoustic artifacts
unrelated to emotion. Both \AA{} and \AB{} use
the same construction procedure within each
corpus and generator. We additionally compare
spliced speech (ESD, naturalistic bona fide
speech, and IndexTTS2) with unspliced speech
(OMG-Emotion, TED-TTS, HybridEmo, and EmoTra), and examine detector scores around known splice
boundaries (\S\ref{sec:q4}).

\paragraph{Duration.}
Within each dataset and authenticity class,
the mean durations of \AA{} and \AB{}
differ by no more than 0.49\,s
(Table~\ref{tab:benchmark}).
These small mean differences suggest that
duration is unlikely to be the primary
driver of the observed score shifts.

\section{Experimental Setup}
\label{sec:setup}

\subsection{Detectors}
\label{sec:detectors}

We evaluate seven speech deepfake detectors spanning
different architectures: RawNet2 \citep{tak2021rawnet2}
(raw waveform), AASIST \citep{jung2022aasist}
(spectro-temporal graph attention), XLS-R-SLS
\citep{zhang2024sls} and XLS-R-Mamba
\citep{xiao2025xlsrmamba} (self-supervised XLS-R
features; \citealp{babu2022xlsr}), ProsDD
\citep{mahapatra2026prosdd} (prosody-aware),
SLIM \citep{zhu2024slim} (style--linguistics
mismatch), and Alethia \citep{marconi}
(foundation model for voice deepfakes).

All detectors are evaluated using checkpoints trained
on ASVspoof 2019 LA \citep{wang2020asvspoof}
to maintain a common training benchmark.
We use publicly released checkpoints for
RawNet2, AASIST, XLS-R-SLS, XLS-R-Mamba,
and ProsDD. 
For SLIM, a suitable ASVspoof 2019 LA
checkpoint was not available, so we
fine-tuned its WavLM-Large backbone
on ASVspoof 2019 LA for five epochs.
For Alethia, we obtained a pretrained
checkpoint from the model authors and
fine-tuned it on ASVspoof 2019 LA.

All audio is converted to 16\,kHz mono.
Each detector scores overlapping windows at its
native input length (3.0--4.2\,s), with 50\%
overlap and the final window aligned to the
end of the utterance.
The utterance-level spoof probability is the
mean of the window-level probabilities.
Full inference settings are provided in
Appendix~\ref{sec:appendix-setup}.

\subsection{Evaluation Protocol}
\label{sec:analyses}

\paragraph{Baseline detection performance.}
We report Equal Error Rate (EER, \%) to assess
overall discrimination between bona fide
and synthetic speech, a widely used metric
in speech deepfake detection
\citep{wang2020asvspoof,wang2024asvspoof5,marconi}.

\paragraph{Emotion-induced score shifts.}
\label{sec:metrics}
To examine how within-utterance emotion changes
affect detector decisions, we compare scores
between emotion-stable (\AA{}) and
emotion-changing (\AB{}) speech separately
for bona fide and synthetic samples.
This distinction allows us to determine
whether emotion changes make synthetic speech
appear more genuine or bona fide speech
appear more synthetic.

Let $p(x)$ be the utterance-level spoof
probability, with higher values indicating
more spoof-like speech.
To capture differences near probability
extremes, we use the logit score
$s(x)=\log\frac{p(x)}{1-p(x)}$,
with $p(x)$ clipped to $[10^{-6},1-10^{-6}]$.
We define
\begin{equation}
\begin{aligned}
\Delta_{\mathrm{SP}} &=
\bar{s}_{\mathrm{SP}}(\AB)
-\bar{s}_{\mathrm{SP}}(\AA),\\
\Delta_{\mathrm{BF}} &=
\bar{s}_{\mathrm{BF}}(\AB)
-\bar{s}_{\mathrm{BF}}(\AA),
\end{aligned}
\end{equation}
where $\bar{s}_{\mathrm{SP}}$ and
$\bar{s}_{\mathrm{BF}}$ are the mean
logit scores of synthetic and bona fide
speech, respectively.
Negative shifts indicate lower spoof scores
for \AB{}, while positive shifts indicate
higher scores.

\begin{figure*}[t]
\centering

\begin{minipage}[c]{0.48\textwidth}
    \centering
    \begingroup
    \scriptsize
    \setlength{\tabcolsep}{2pt}
    \renewcommand{\arraystretch}{0.95}
    \resizebox{\linewidth}{!}{%
        \begin{tabular}{@{}l rr rr rr rr@{}}
\toprule
 & \multicolumn{2}{c}{ESD} & \multicolumn{2}{c}{OMG-Emotion} & \multicolumn{2}{c}{IEMOCAP} & \multicolumn{2}{c}{MSP-Podcast} \\
 & \multicolumn{2}{c}{\emph{controlled}} & \multicolumn{2}{c}{\emph{in-the-wild}} & \multicolumn{4}{c}{\emph{naturalistic}} \\
\cmidrule(lr){2-3}\cmidrule(lr){4-5}\cmidrule(lr){6-7}\cmidrule(l){8-9}
Detector & $\Delta_{\text{SP}}$ & $\Delta_{\text{BF}}$ & $\Delta_{\text{SP}}$ & $\Delta_{\text{BF}}$ & $\Delta_{\text{SP}}$ & $\Delta_{\text{BF}}$ & $\Delta_{\text{SP}}$ & $\Delta_{\text{BF}}$ \\
\midrule
RawNet2 & +0.01 & \textminus{}0.28 & \textbf{\textminus{}3.25} & +0.10 & \cellcolor{gray!15}+0.09 & \cellcolor{gray!15}\textbf{+1.54} & \cellcolor{gray!15}+0.03 & \cellcolor{gray!15}+0.13 \\
AASIST & \textbf{\textminus{}0.22} & \textbf{\textminus{}0.50} & \textbf{\textminus{}1.64} & \textminus{}0.15 & \cellcolor{gray!15}\textbf{+0.25} & \cellcolor{gray!15}\textbf{+1.66} & \cellcolor{gray!15}\textminus{}0.02 & \cellcolor{gray!15}+0.23 \\
XLS-R-SLS & \textbf{\textminus{}0.85} & \textbf{\textminus{}0.42} & \textbf{\textminus{}2.64} & \textminus{}0.03 & \textbf{+0.49} & \textbf{\textminus{}0.60} & \cellcolor{gray!15}\textminus{}0.04 & \cellcolor{gray!15}0.00 \\
XLS-R-Mamba & \textbf{\textminus{}0.67} & \textminus{}0.08 & \textbf{\textminus{}2.47} & +0.30 & \cellcolor{gray!15}\textbf{+0.39} & \cellcolor{gray!15}\textbf{\textminus{}0.75} & \cellcolor{gray!15}\textminus{}0.02 & \cellcolor{gray!15}+0.11 \\
ProsDD & \textbf{\textminus{}0.37} & \textbf{\textminus{}0.56} & \cellcolor{gray!15}\textbf{\textminus{}2.02} & \cellcolor{gray!15}\textminus{}0.57 & \cellcolor{gray!15}\textbf{+0.57} & \cellcolor{gray!15}\textminus{}0.27 & \cellcolor{gray!15}+0.14 & \cellcolor{gray!15}\textbf{+0.83} \\
SLIM & +0.02 & \textbf{+0.23} & \cellcolor{gray!15}\textbf{\textminus{}0.46} & \cellcolor{gray!15}+0.02 & \cellcolor{gray!15}\textbf{+0.32} & \cellcolor{gray!15}+0.14 & \cellcolor{gray!15}0.00 & \cellcolor{gray!15}\textbf{+0.39} \\
Alethia & \textbf{\textminus{}0.52} & +0.23 & \textbf{\textminus{}2.03} & \textminus{}0.35 & \cellcolor{gray!15}\textbf{+0.99} & \cellcolor{gray!15}\textbf{+0.79} & \cellcolor{gray!15}+0.06 & \cellcolor{gray!15}\textbf{+0.92} \\
\bottomrule
\end{tabular}

    }
    \endgroup
\end{minipage}
\hfill
\begin{minipage}[c]{0.51\textwidth}
    \centering
    \includegraphics[
        width=\linewidth,
        trim=8 5 8 5,
        clip
    ]{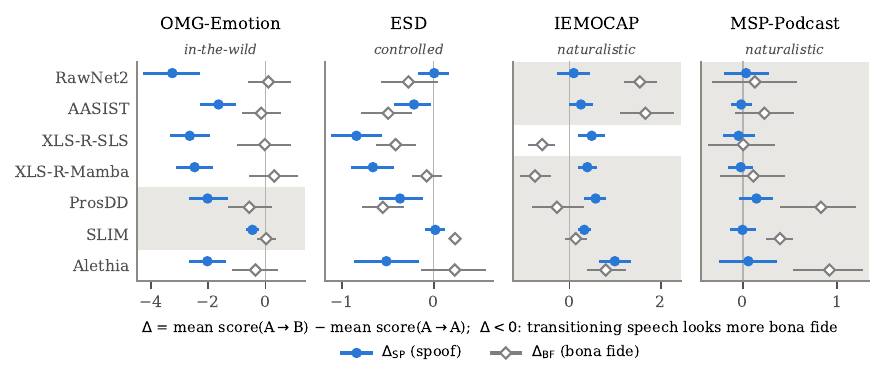}
\end{minipage}

\vspace{-4pt}
\caption{
Detector score shifts from emotion-stable
(\AA{}) to emotion-changing (\AB{}) speech.
\emph{Left:} mean logit score differences for
synthetic ($\Delta_{\mathrm{SP}}$) and bona fide
($\Delta_{\mathrm{BF}}$) speech; bold indicates
95\% CIs excluding zero.
\emph{Right:} corresponding score shifts
with 95\% CIs (filled: synthetic; hollow:
bona fide).
Negative values indicate lower spoof scores
for \AB{}.
Grey indicates pooled EER above 50\%
(Table~\ref{tab:auc}).
Additional results are in
Appendix Table~\ref{tab:full}.
}
\label{fig:shift}
\vspace{-7pt}
\end{figure*}

For ESD, we compute differences within
matched \AA{}/\AB{} pairs and average them.
For the other datasets, where \AA{} and \AB{}
use separate source utterances, we subtract
the mean score of all \AA{} samples from
that of all \AB{} samples within each
authenticity class.
These score shifts are reported in
Figure~\ref{fig:shift}.

We additionally examine
$\Delta_{\mathrm{SP}}-\Delta_{\mathrm{BF}}$
to assess changes in the mean score
separation between bona fide and synthetic
speech (Appendix Table~\ref{tab:full}).

\paragraph{Detection errors.}
We compare EER between \AA{} and \AB{}
to assess changes in overall detection
performance.
However, EER allows the decision threshold
to vary between conditions, potentially
obscuring changes in errors at a fixed
operating point.
We therefore calibrate a threshold at the
EER point on \AA{} speech and apply it
unchanged to \AB{}.
We report changes in false-negative rate
($\Delta$FNR; synthetic speech incorrectly flagged as bonafide) and
false-positive rate ($\Delta$FPR; bona fide
speech incorrectly flagged as synthetic),
in percentage points (\AB{} minus \AA{}).

\paragraph{Emotion and splicing analyses.}
We examine individual emotion transitions
in ESD and arousal changes in naturalistic
speech to characterize how detector responses
vary with emotional trajectories
(\S\ref{sec:emotion}).
We also compare spliced and jointly generated
speech and analyze scores around known splice
boundaries to examine potential effects of
concatenation artifacts (\S\ref{sec:q4}).

\paragraph{Statistical uncertainty.}
We estimate uncertainty in detector score shifts
using 1,000 bootstrap resamples of utterances,
separately within each dataset and
authenticity class.
For datasets with speaker IDs, we resample
speakers with replacement, retaining their
utterances and \AA{}/\AB{} assignments.
We recompute score shifts for each resample
and report 95\% bootstrap confidence intervals.
A shift is considered statistically significant
when its confidence interval excludes zero.



\section{Results}
\label{sec:results}

Figure~\ref{fig:shift} summarizes how
within-utterance emotion changes affect
detector scores across four datasets.
We examine synthetic speech, followed
by bona fide speech, before investigating
emotion-specific responses and
alternative explanations.

Baseline detection performance varies
substantially across datasets.
On OMG-Emotion, SLIM and ProsDD have EERs
above 50\%, while EERs on ESD range from
13.0\% to 43.2\%.
On IEMOCAP and MSP-Podcast, most detectors
have EERs above 50\%.
Full baseline results are provided in
Appendix Table~\ref{tab:auc}.
These baseline differences provide context
for interpreting the score shifts below.


\subsection{How Do Emotion Transitions Affect Synthetic Speech Detection?}
\label{sec:q1}

\paragraph{Categorical emotion transitions.}
On OMG-Emotion, all seven detectors assign
significantly lower spoof scores to
emotion-changing (\AB{}) than to
emotion-stable (\AA{}) synthetic speech,
while bona fide scores remain largely unchanged
(Figure~\ref{fig:shift}).
This indicates that emotional transitions make
synthetic speech appear more genuine to the
detectors, without similarly affecting their
judgments of real speech.
At thresholds fixed on \AA{}, spoof miss rates
increase by 9.7--25.3 percentage points
(Appendix Table~\ref{tab:full}).
Thus, detectors miss more transitioning deepfakes
at the same threshold.

A similar pattern appears on ESD, where
\AA{} and \AB{} share the same speaker and
text, allowing a more controlled comparison.
Five of the seven detectors assign
lower spoof scores to \AB{}, making transitioning
synthetic speech appear more genuine.
For XLS-R-SLS and XLS-R-Mamba, spoof miss rates
increase by 6.8 and 4.3 percentage points,
respectively, while EER changes are small
and not statistically significant
($+1.1$ and $+1.6$ points).
This shows that EER alone may not capture
the increased risk of missed
transitioning deepfakes.

\paragraph{Continuous VAD-based transitions.}
A different pattern emerges on IEMOCAP
and MSP-Podcast, where emotional
transitions are selected using continuous
valence, arousal, and dominance (VAD) annotations.
On IEMOCAP, six of the seven detectors assign
significantly higher spoof scores to \AB{},
making transitioning synthetic speech appear
more fake rather than more genuine.
On MSP-Podcast, spoof scores remain nearly
unchanged across all seven detectors, although
ProsDD, SLIM, and Alethia assign significantly
higher spoof scores to transitioning bona fide
speech, as discussed in the next section.

Unlike categorical transitions in ESD
and OMG-Emotion, VAD-based transitions do
not consistently make synthetic speech harder
to detect. These differences may reflect
differences in emotion conditioning, synthesis
methods, or source characteristics.


\subsection{How Do Detectors Respond to Genuine
Emotion Transitions?}
\label{sec:q2}

Emotion transitions also change how detectors
score bona fide speech, but the direction
varies across models and corpora.

On ESD, SLIM assigns significantly higher
spoof scores to transitioning bona fide
speech ($\Delta_{\mathrm{BF}}=+0.23$).
A similar effect appears on IEMOCAP, where
AASIST and RawNet2 increase bona fide spoof
scores by $+1.66$ and $+1.54$, respectively.
At fixed thresholds, this corresponds to
18.6--24.4 percentage points of more false alarms.

On MSP-Podcast, ProsDD, SLIM, and Alethia
also assign significantly higher spoof scores
to genuine transitions ($+0.83$, $+0.39$,
and $+0.92$, respectively).
To assess whether these effects persist after
accounting for differences in speaker,
duration, and regional VAD, we perform an
adjusted analysis.
SLIM and Alethia continue to show positive
score shifts on both IEMOCAP and MSP-Podcast
(Appendix Table~\ref{tab:nat}).

In contrast, the XLS-R detectors assign lower
spoof scores to transitioning bona fide speech
on IEMOCAP. This effect persists after adjustment, with
score shifts of $-0.60$ for XLS-R-SLS and
$-0.82$ for XLS-R-Mamba
(Appendix Table~\ref{tab:nat}).
For both XLS-R detectors, synthetic speech
shifts in the opposite direction, increasing
the mean score separation between bona fide
and synthetic speech (Figure~\ref{fig:shift}).
EER decreases from 48.6\% to 39.7\%
for XLS-R-SLS, and from 59.0\% to 52.5\%
for XLS-R-Mamba, although only the former
change is statistically significant
(Appendix Table~\ref{tab:full}).

Overall, genuine emotion transitions affect
detectors differently. Some assign higher
spoof scores to transitioning real speech,
while others assign lower scores.
These differences show that the effect of
emotional change depends on the detector
and the speech conditions. For deployment in the wild, detector algorithms need a systematic redesign to be robust to emotional change within an utterance.


\begin{figure*}[!t]
\centering
\includegraphics[
    width=\textwidth,
    height=5.5cm,
    keepaspectratio
]{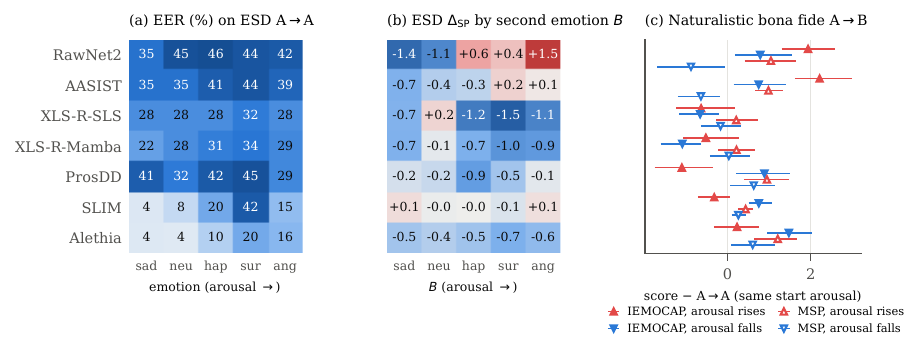}
\vspace{-5pt}
\caption{
Emotion-specific detector responses.
(a) EER on emotion-stable (\AA{}) ESD speech
by emotion.
(b) ESD spoof score shifts by destination
emotion.
(c) Bona fide spoof score shifts for
transitions with increasing or decreasing
arousal, relative to stable speech with
similar starting arousal.
Positive values indicate higher spoof scores
for transitioning bona fide speech.
Error bars show 95\% bootstrap CIs.
}
\label{fig:emotion}
\vspace{-8pt}
\end{figure*}

\subsection{Why Do Detectors Respond Differently?}
\label{sec:q3}
\label{sec:emotion}

The preceding results show that emotional
transitions affect synthetic and genuine speech
differently across detectors.
We further investigate whether these effects
depend on the specific emotion trajectory.
We use ESD to examine destination emotions
under controlled conditions, and IEMOCAP
and MSP-Podcast to study changes in arousal
using continuous VAD annotations.

\paragraph{Sensitivity to individual emotions.}
Even without emotion transitions, detection
performance varies across emotions
(Figure~\ref{fig:emotion}a).
Surprise generally yields higher EERs,
while sad and neutral speech yield lower EERs.
This is especially pronounced for SLIM,
whose EER rises from 4\% on sad speech to
42\% on surprise, and Alethia, whose EER
rises from 4\% to 20\%.
However, this pattern is not universal:
ProsDD performs best on angry speech,
while XLS-R detectors show smaller
differences across emotions.

\paragraph{The destination emotion matters.}
On ESD, the XLS-R detectors consistently
assign lower spoof scores to transitions
ending in non-neutral emotions
(Figure~\ref{fig:emotion}b).
For XLS-R-SLS, shifts range from $-0.7$
to $-1.5$, compared with $+0.2$ for neutral.

RawNet2 and AASIST show a different pattern:
spoof scores tend to decrease for sad and
neutral endpoints but increase for more
arousing emotions.
For RawNet2, shifts range from $-1.4$
for sadness to $+1.5$ for anger.
These opposing shifts largely cancel,
explaining its near-zero overall score
change on ESD ($+0.01$).

Among other detectors, ProsDD shows
larger reductions for happy and surprise,
Alethia shows negative shifts across all
endpoints, and SLIM remains close to zero
across destination emotions.

\paragraph{The direction of arousal change matters.}
Figure~\ref{fig:emotion}c shows that detector
score shifts differ between increasing and
decreasing arousal, with patterns varying
across detectors and datasets.
We distinguish directions using
the arousal difference between the
second and first regions.

RawNet2 and AASIST show particularly
consistent responses across the two
naturalistic datasets.
On IEMOCAP, compared with emotion-stable
speech with similar starting arousal,
their spoof scores increase by $+1.9$
and $+2.2$ logits, respectively, when
arousal increases.
When arousal decreases, the increases
are smaller ($+0.8$ for both).
A similar pattern appears on MSP-Podcast.

We further examine the relationship between
arousal and detector scores within individual
utterances (Appendix~\ref{sec:appendix-nat}).
Only RawNet2 and AASIST show consistently
positive associations between arousal changes
and spoof scores across both datasets.
Together, these findings reveal broader
detector-specific sensitivities to arousal
direction, with RawNet2 and AASIST showing
the most consistent tendency to assign
higher spoof scores to increasing arousal
in bona fide speech.


\subsection{Are the Effects Explained by
Emotion Identity or Splicing?}
\label{sec:q4}

\paragraph{Emotion identity.}
As shown above, detector
performance varies across individual emotions.
We therefore examine whether the score shifts
between \AA{} and \AB{} are caused
by the emotions present, rather than by
the transition itself.

On ESD, all 20 ordered emotion transitions
are balanced, with each emotion appearing
equally often in both regions.
Despite this balance, five detectors assign
significantly lower spoof scores to \AB{}
(Figure~\ref{fig:shift}), suggesting that
differences in emotion frequency alone
cannot explain the effect.

On OMG-Emotion, we further account for the
starting and ending emotions of each transition.
For an \AB{} spoof, we compare its score with
the average scores of stable-emotion spoofs
expressing emotion $A$ (\AA{}) and emotion
$B$ ($B\to B$), using samples from the same
generator. These controls are grouped by
emotion rather than matched by text.
After this comparison, all seven detectors
still assign significantly lower scores
to transitioning spoofs ($-0.41$ to $-3.42$). These results suggest that the lower spoof
scores cannot be explained solely by the
emotions present in the speech.

\paragraph{Splicing artifacts.}
Some benchmark utterances are constructed
by concatenating two speech segments,
which may introduce acoustic artifacts
unrelated to emotion transitions.
We examine whether these artifacts could
explain the observed score shifts in both
synthetic and bona fide speech.

For synthetic speech, IndexTTS2 generates
and concatenates two segments, whereas
TED-TTS and HybridEmo generate complete
utterances without splicing.
On OMG-Emotion, all three generators show
significantly lower spoof scores for \AB{}
than for \AA{} across nearly all detectors
(7/7 for IndexTTS2, 6/7 for TED-TTS,
and 7/7 for HybridEmo).
On ESD, the jointly generated TED-TTS
shows significant score reductions for
more detectors than the spliced IndexTTS2
(5/7 versus 3/7).
These results
(Appendix Figure~\ref{fig:splice}a)
show that spoof score reductions
do not require splicing.

For bona fide speech, ESD uses the same
concatenation procedure for \AA{} and \AB{},
yet score differences remain.
We further examine IEMOCAP and MSP-Podcast
by comparing scores from windows containing
the splice boundary with other windows
in the same utterance.
Alethia, SLIM, and ProsDD assign higher
spoof scores to splice-boundary windows,
while unspliced EmoTra spoofs show little
change at the corresponding position.
However, the boundary effect is similar
for bona fide \AA{} and \AB{}.
For example, Alethia shows increases of
$+4.5$ and $+4.7$ logits, respectively,
on MSP-Podcast (Appendix Figure~\ref{fig:splice}). Thus, splicing affects local detector scores,
but has similar effects on emotion-stable
(\AA{}) and emotion-changing (\AB{}) speech
and thus cannot fully explain the
differences between conditions.


\section{Discussion}

\paragraph{Emotional variation can reduce spoof detectability.}
A notable finding is that emotion changes
can make synthetic speech appear more genuine
to deepfake detectors. On ESD and OMG-Emotion,
several detectors assign lower spoof scores
to emotion-changing synthetic speech, even
though the speech remains synthetic.
This suggests that some emotion-related
acoustic patterns may overlap with features
associated with bona fide speech.
These score shifts are also associated with
increased spoof miss rates at fixed detection
thresholds, highlighting a potential risk
in deployment.

\paragraph{The effect depends on how emotions
change and how they are synthesized.}
The contrast between categorical and
continuous VAD-based transitions is
particularly informative. Strong spoof
score reductions occur in ESD and
OMG-Emotion, whereas naturalistic corpora
show smaller or opposite shifts.
Moreover, the same XLS-R detectors that
assign lower scores to transitioning spoofs
on ESD and OMG-Emotion can show improved
class separation on IEMOCAP. One possible explanation is that
emotion-related acoustic cues may have
different distributions in bona fide
and synthetic speech, leading to
different detector responses.
Consistent with this possibility, our
emotion-specific analyses reveal that
XLS-R responses vary with destination
emotion, while RawNet2 and AASIST show
associations with arousal direction.
These observations suggest that the effect
of a transition depends on both the detector
and how the emotional change is realized.

\paragraph{Implications for deepfake detection.}
Most emotion-aware evaluations characterize
speech using utterance-level emotion labels,
which do not capture emotion changes occurring
within an utterance. Our findings show that
these temporal changes can affect detector
scores and error rates in ways that are
not fully explained by static emotion
identity or splicing artifacts.
Evaluating both categorical transitions
and continuous VAD trajectories
provides a more complete view of detector
behavior. In addition, fixed-threshold
evaluation is important because increases
in missed spoofs or false alarms may be
less apparent from EER alone. Future
detectors would benefit from training and
evaluation data that include emotional
transitions in both bona fide and
synthetic speech, rather than treating
emotion as a static utterance-level property.

\section{Conclusion}

We introduced \bench{}, a benchmark
evaluating the effects of categorical and
continuous VAD-based emotion transitions
on speech deepfake detection.
We find that transitions can increase
missed spoofs even with small EER changes,
while genuine transitions also shift scores
depending on the detector and emotion trajectory.
These effects are not fully explained by
static emotion identity or splicing artifacts,
highlighting within-utterance emotion dynamics
as an important evaluation dimension for
speech deepfake detection.

\section*{Limitations}

Our evaluation covers four source datasets,
multiple synthesis methods, and seven
deepfake detectors, providing complementary
controlled and naturalistic settings.
As with any benchmark, the observed effects
reflect the speech conditions and synthesis
systems included in the evaluation.
The categorical and VAD-based settings also
differ in their generation methods, so their
individual contributions cannot be isolated.
In the naturalistic datasets, bona fide and
synthetic speech differ in their construction
procedures, which may affect cross-class
detection metrics. Our primary analyses
compare emotion-stable and emotion-changing
speech within each authenticity class,
reducing this potential confound.
Future work can extend \bench{} to additional
languages, synthesis systems, and emotion
trajectories to further assess the
generalizability of these findings.

\section*{Ethics Statement}

This work investigates how emotional changes
in speech affect the reliability of audio
deepfake detection systems. The benchmark
is intended to support the development
and evaluation of more robust detection
methods, particularly for emotionally
expressive speech.

Our experiments use established speech
corpora and synthetic audio generated for
research evaluation. We follow applicable
dataset access and usage conditions and
report results at the aggregate level.
The study does not involve deploying
synthetic speech for impersonation or
deception. We hope the findings will
contribute to more reliable and responsible
audio authenticity assessment.

\bibliography{custom}

\newpage

\appendix


\section{Dataset Licenses and Usage Conditions}
\label{sec:dataset-licenses}

We use four existing speech corpora under
their respective licenses and usage agreements.
The necessary permissions have been obtained
for the use and distribution of the benchmark
artifacts released in this work.

\paragraph{ESD.}
The Emotional Speech Dataset (ESD) is
distributed for research use under
a dataset-specific license agreement,
separate from the repository's MIT license.

\paragraph{OMG-Emotion.}
OMG-Emotion is distributed under
CC BY-NC-SA 3.0 DE, subject to additional
research-use and redistribution conditions.

\paragraph{IEMOCAP.}
IEMOCAP is provided by USC SAIL under
a non-transferable research license.
Its use and redistribution are governed
by the applicable licensing agreement
and permissions.

\paragraph{MSP-Podcast.}
MSP-Podcast is distributed by UT Dallas
under academic and commercial licensing
agreements. Its use and redistribution
are subject to the applicable agreement
and source recording rights.

\section{Benchmark Details}
\label{sec:appendix-data}

\paragraph{Benchmark distribution.}
\bench{} includes both bona fide and
synthetically generated speech derived
from these resources. The original
datasets retain their respective
licensing and usage conditions.
Any redistribution of benchmark audio
must comply with the applicable
permissions for the underlying data
and generated artifacts.

\paragraph{Selection of naturalistic emotion changes.}
For IEMOCAP and MSP-Podcast, we characterize
emotional differences between two speech
regions using the Euclidean distance between
their normalized valence, arousal, and
dominance (VAD) annotations.
Each VAD dimension is normalized to $[0,1]$.

We construct emotion-stable (\AA{}) speech
from regions sharing the same categorical
emotion with small VAD distances, and
emotion-changing (\AB{}) speech from regions
with different categorical emotions and
large VAD distances.
Table~\ref{tab:vad_selection} reports the
selection thresholds and the distributions
of VAD distances in the final subsets.
The selected \AA{} and \AB{} conditions
have non-overlapping VAD distance ranges
in both datasets.
\begin{table}[t]
\centering
\small
\setlength{\tabcolsep}{4pt}
\resizebox{\columnwidth}{!}{%
\begin{tabular}{@{}llccccc@{}}
\toprule
 & & \multicolumn{2}{c}{Selection threshold}
 & \multicolumn{3}{c}{Selected $d_{\text{VAD}}$} \\
\cmidrule(lr){3-4}
\cmidrule(l){5-7}
Dataset & Cond. & \AA{} $\leq$ & \AB{} $\geq$
& Mean & Median & Range \\
\midrule
IEMOCAP & \AA{}
& \multirow{2}{*}{0.1816}
& \multirow{2}{*}{0.6305}
& 0.033 & 0.000 & 0.000--0.125 \\
& \AB{} & & &
0.958 & 0.972 & 0.839--1.087 \\
MSP-Podcast & \AA{}
& \multirow{2}{*}{0.1414}
& \multirow{2}{*}{0.5509}
& 0.076 & 0.075 & 0.000--0.141 \\
& \AB{} & & &
0.690 & 0.626 & 0.551--1.397 \\
\bottomrule
\end{tabular}}
\caption{
VAD-based selection of naturalistic speech.
$d_{\mathrm{VAD}}$ denotes the Euclidean
distance between the normalized VAD vectors
of two speech regions.
Selection thresholds define candidate pools;
the final selected samples show stronger
separation, with non-overlapping \AA{}
and \AB{} distance ranges.
}
\label{tab:vad_selection}
\end{table}

\section{Full Detection Results}
\label{sec:appendix-results}

\paragraph{Baseline detection performance.}
Table~\ref{tab:auc} reports the pooled
equal error rate (EER) and area under
the ROC curve (AUC) for each detector
and dataset, combining \AA{} and \AB{} speech.
Grey shading identifies detector--dataset
combinations with pooled EER above 50\%,
indicating inverted score rankings.
These baseline results provide context
for interpreting subsequent score shifts
and detection errors.

\begin{table}[t]
\centering
\small
\setlength{\tabcolsep}{4pt}
\resizebox{\columnwidth}{!}{%
\begin{tabular}{@{}lcccc@{}}
\toprule
Detector & ESD & OMG & IEMOCAP & MSP \\
\midrule
\multicolumn{5}{@{}l}{\emph{EER (\%), lower is better}} \\
RawNet2 & 43.2 & 35.9 & \cellcolor{gray!12}\textcolor{gray}{62.4} & \cellcolor{gray!12}\textcolor{gray}{81.4} \\
AASIST & 39.4 & 35.7 & \cellcolor{gray!12}\textcolor{gray}{58.9} & \cellcolor{gray!12}\textcolor{gray}{86.3} \\
XLS-R-SLS & 30.4 & 41.4 & 44.2 & \cellcolor{gray!12}\textcolor{gray}{50.9} \\
XLS-R-Mamba & 29.1 & 45.7 & \cellcolor{gray!12}\textcolor{gray}{55.6} & \cellcolor{gray!12}\textcolor{gray}{56.5} \\
ProsDD & 36.6 & \cellcolor{gray!12}\textcolor{gray}{56.2} & \cellcolor{gray!12}\textcolor{gray}{56.8} & \cellcolor{gray!12}\textcolor{gray}{53.5} \\
SLIM & 20.5 & \cellcolor{gray!12}\textcolor{gray}{73.0} & \cellcolor{gray!12}\textcolor{gray}{84.4} & \cellcolor{gray!12}\textcolor{gray}{76.1} \\
Alethia & 13.0 & 42.3 & \cellcolor{gray!12}\textcolor{gray}{65.6} & \cellcolor{gray!12}\textcolor{gray}{59.3} \\
\midrule
\multicolumn{5}{@{}l}{\emph{AUC, higher is better}} \\
RawNet2 & 0.60 & 0.66 & \cellcolor{gray!12}\textcolor{gray}{0.27} & \cellcolor{gray!12}\textcolor{gray}{0.11} \\
AASIST & 0.65 & 0.69 & \cellcolor{gray!12}\textcolor{gray}{0.34} & \cellcolor{gray!12}\textcolor{gray}{0.07} \\
XLS-R-SLS & 0.77 & 0.63 & 0.58 & \cellcolor{gray!12}\textcolor{gray}{0.48} \\
XLS-R-Mamba & 0.78 & 0.58 & \cellcolor{gray!12}\textcolor{gray}{0.45} & \cellcolor{gray!12}\textcolor{gray}{0.40} \\
ProsDD & 0.69 & \cellcolor{gray!12}\textcolor{gray}{0.41} & \cellcolor{gray!12}\textcolor{gray}{0.40} & \cellcolor{gray!12}\textcolor{gray}{0.47} \\
SLIM & 0.87 & \cellcolor{gray!12}\textcolor{gray}{0.22} & \cellcolor{gray!12}\textcolor{gray}{0.13} & \cellcolor{gray!12}\textcolor{gray}{0.22} \\
Alethia & 0.95 & 0.62 & \cellcolor{gray!12}\textcolor{gray}{0.32} & \cellcolor{gray!12}\textcolor{gray}{0.42} \\
\bottomrule
\end{tabular}}
\caption{Baseline detection per corpus, pooling \AA{} and \AB{}. Grey: EER above 50\% (equivalently, AUC below 0.5 for every cell here), i.e., the detector ranks bona fide speech as more spoof-like than spoofs in most bona fide--spoof pairs.}
\label{tab:auc}
\end{table}

\begin{figure}[!htbp]
\centering
\includegraphics[width=\columnwidth]
{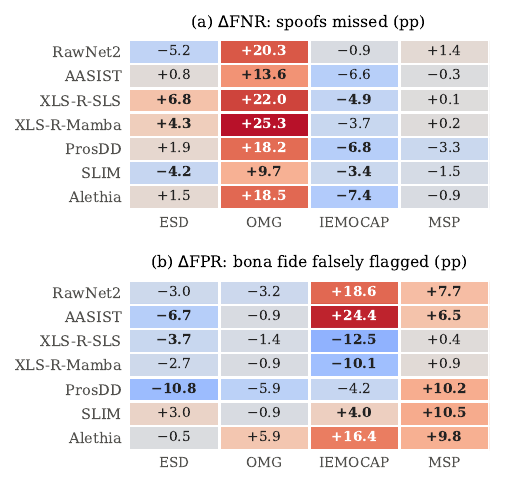}
\caption{
Changes in detection errors from \AA{}
to \AB{} at thresholds calibrated
on \AA{} speech.
Values are in percentage points.
Positive changes indicate higher error rates
for emotion-changing speech.
Bold indicates 95\% bootstrap confidence
intervals excluding zero.
}
\label{fig:cost-cells}
\end{figure}

\paragraph{Emotion-induced detection errors.}
Figure~\ref{fig:cost-cells} summarizes changes
in false-negative and false-positive rates
between emotion-stable and emotion-changing
speech, using thresholds calibrated at
the EER operating point on \AA{}.
The same threshold is applied to \AB{},
allowing us to examine changes in detection
errors without recalibrating for emotional
transitions.

\paragraph{Complete detector comparisons.}
Table~\ref{tab:full} provides the complete
results for all seven detectors across
the four datasets.
For each detector, we report the mean
score shifts for synthetic
($\Delta_{\mathrm{SP}}$) and bona fide
($\Delta_{\mathrm{BF}}$) speech, their
difference
($\Delta_{\mathrm{SP}}-\Delta_{\mathrm{BF}}$),
EER under \AA{} and \AB{}, and changes
in false-negative and false-positive rates
at the fixed \AA{} threshold.

The difference
$\Delta_{\mathrm{SP}}-\Delta_{\mathrm{BF}}$
describes the change in signed mean score
separation between the two authenticity
classes. It complements, rather than
replaces, the EER and fixed-threshold
error measurements.

\begin{table*}[!htbp]
\centering
\small
\setlength{\tabcolsep}{5pt}
\begin{tabular}{@{}cl rrr c rr@{}}
\toprule
 & & \multicolumn{3}{c}{Score difference (logit)} & EER (\%) & \multicolumn{2}{c}{Fixed threshold (pp)} \\
\cmidrule(lr){3-5}\cmidrule(lr){6-6}\cmidrule(l){7-8}
 & Detector & $\Delta_{\text{SP}}$ & $\Delta_{\text{BF}}$ & $\Delta_{\text{SP}}{-}\Delta_{\text{BF}}$ & \AA{} $\to$ \AB{} & $\Delta$FNR & $\Delta$FPR \\
\midrule
\multirow{7}{*}{\rotatebox{90}{ESD}} & RawNet2 & +0.01 & \textminus{}0.28 & +0.28 & \textbf{45.2 $\to$ 40.8} & \textminus{}5.2 & \textminus{}3.0 \\
 & AASIST & \textbf{\textminus{}0.22} & \textbf{\textminus{}0.50} & \textbf{+0.28} & 41.0 $\to$ 38.2 & +0.8 & \textbf{\textminus{}6.7} \\
 & XLS-R-SLS & \textbf{\textminus{}0.85} & \textbf{\textminus{}0.42} & \textbf{\textminus{}0.43} & 28.7 $\to$ 29.8 & \textbf{+6.8} & \textbf{\textminus{}3.7} \\
 & XLS-R-Mamba & \textbf{\textminus{}0.67} & \textminus{}0.08 & \textbf{\textminus{}0.59} & 29.2 $\to$ 30.8 & \textbf{+4.3} & \textminus{}2.7 \\
 & ProsDD & \textbf{\textminus{}0.37} & \textbf{\textminus{}0.56} & +0.19 & \textbf{39.2 $\to$ 34.5} & +1.9 & \textbf{\textminus{}10.8} \\
 & SLIM & +0.02 & \textbf{+0.23} & \textbf{\textminus{}0.22} & 20.5 $\to$ 20.2 & \textbf{\textminus{}4.2} & +3.0 \\
 & Alethia & \textbf{\textminus{}0.52} & +0.23 & \textbf{\textminus{}0.75} & 12.8 $\to$ 13.4 & +1.5 & \textminus{}0.5 \\
\midrule
\multirow{7}{*}{\rotatebox{90}{OMG}} & RawNet2 & \textbf{\textminus{}3.25} & +0.10 & \textbf{\textminus{}3.35} & \textbf{31.8 $\to$ 41.4} & \textbf{+20.3} & \textminus{}3.2 \\
 & AASIST & \textbf{\textminus{}1.64} & \textminus{}0.15 & \textbf{\textminus{}1.48} & \textbf{32.7 $\to$ 39.9} & \textbf{+13.6} & \textminus{}0.9 \\
 & XLS-R-SLS & \textbf{\textminus{}2.64} & \textminus{}0.03 & \textbf{\textminus{}2.61} & \textbf{35.9 $\to$ 45.5} & \textbf{+22.0} & \textminus{}1.4 \\
 & XLS-R-Mamba & \textbf{\textminus{}2.47} & +0.30 & \textbf{\textminus{}2.77} & \textbf{38.7 $\to$ 53.2} & \textbf{+25.3} & \textminus{}0.9 \\
 & \cellcolor{gray!12}\textcolor{gray}{ProsDD$^\dagger$} & \cellcolor{gray!12}\textbf{\textminus{}2.02} & \cellcolor{gray!12}\textminus{}0.57 & \cellcolor{gray!12}\textbf{\textminus{}1.45} & \cellcolor{gray!12}\textbf{54.5 $\to$ 60.9} & \cellcolor{gray!12}\textbf{+18.2} & \cellcolor{gray!12}\textminus{}5.9 \\
 & \cellcolor{gray!12}\textcolor{gray}{SLIM$^\dagger$} & \cellcolor{gray!12}\textbf{\textminus{}0.46} & \cellcolor{gray!12}+0.02 & \cellcolor{gray!12}\textbf{\textminus{}0.48} & \cellcolor{gray!12}\textbf{70.1 $\to$ 75.9} & \cellcolor{gray!12}\textbf{+9.7} & \cellcolor{gray!12}\textminus{}0.9 \\
 & Alethia & \textbf{\textminus{}2.03} & \textminus{}0.35 & \textbf{\textminus{}1.67} & \textbf{36.4 $\to$ 46.4} & \textbf{+18.5} & +5.9 \\
\midrule
\multirow{7}{*}{\rotatebox{90}{IEMOCAP}} & \cellcolor{gray!12}\textcolor{gray}{RawNet2$^\dagger$} & \cellcolor{gray!12}+0.09 & \cellcolor{gray!12}\textbf{+1.54} & \cellcolor{gray!12}\textbf{\textminus{}1.45} & \cellcolor{gray!12}\textbf{60.5 $\to$ 65.5} & \cellcolor{gray!12}\textminus{}0.9 & \cellcolor{gray!12}\textbf{+18.6} \\
 & \cellcolor{gray!12}\textcolor{gray}{AASIST$^\dagger$} & \cellcolor{gray!12}\textbf{+0.25} & \cellcolor{gray!12}\textbf{+1.66} & \cellcolor{gray!12}\textbf{\textminus{}1.41} & \cellcolor{gray!12}\textbf{53.2 $\to$ 67.6} & \cellcolor{gray!12}\textminus{}6.6 & \cellcolor{gray!12}\textbf{+24.4} \\
 & XLS-R-SLS & \textbf{+0.49} & \textbf{\textminus{}0.60} & \textbf{+1.08} & \textbf{48.6 $\to$ 39.7} & \textbf{\textminus{}4.9} & \textbf{\textminus{}12.5} \\
 & \cellcolor{gray!12}\textcolor{gray}{XLS-R-Mamba$^\dagger$} & \cellcolor{gray!12}\textbf{+0.39} & \cellcolor{gray!12}\textbf{\textminus{}0.75} & \cellcolor{gray!12}\textbf{+1.14} & \cellcolor{gray!12}\textbf{59.0 $\to$ 52.5} & \cellcolor{gray!12}\textminus{}3.7 & \cellcolor{gray!12}\textbf{\textminus{}10.1} \\
 & \cellcolor{gray!12}\textcolor{gray}{ProsDD$^\dagger$} & \cellcolor{gray!12}\textbf{+0.57} & \cellcolor{gray!12}\textminus{}0.27 & \cellcolor{gray!12}\textbf{+0.84} & \cellcolor{gray!12}\textbf{59.4 $\to$ 54.1} & \cellcolor{gray!12}\textbf{\textminus{}6.8} & \cellcolor{gray!12}\textminus{}4.2 \\
 & \cellcolor{gray!12}\textcolor{gray}{SLIM$^\dagger$} & \cellcolor{gray!12}\textbf{+0.32} & \cellcolor{gray!12}+0.14 & \cellcolor{gray!12}+0.19 & \cellcolor{gray!12}85.1 $\to$ 84.2 & \cellcolor{gray!12}\textbf{\textminus{}3.4} & \cellcolor{gray!12}\textbf{+4.0} \\
 & \cellcolor{gray!12}\textcolor{gray}{Alethia$^\dagger$} & \cellcolor{gray!12}\textbf{+0.99} & \cellcolor{gray!12}\textbf{+0.79} & \cellcolor{gray!12}+0.20 & \cellcolor{gray!12}63.4 $\to$ 66.9 & \cellcolor{gray!12}\textbf{\textminus{}7.4} & \cellcolor{gray!12}\textbf{+16.4} \\
\midrule
\multirow{7}{*}{\rotatebox{90}{MSP}} & \cellcolor{gray!12}\textcolor{gray}{RawNet2$^\dagger$} & \cellcolor{gray!12}+0.03 & \cellcolor{gray!12}+0.13 & \cellcolor{gray!12}\textminus{}0.09 & \cellcolor{gray!12}\textbf{78.5 $\to$ 84.5} & \cellcolor{gray!12}+1.4 & \cellcolor{gray!12}\textbf{+7.7} \\
 & \cellcolor{gray!12}\textcolor{gray}{AASIST$^\dagger$} & \cellcolor{gray!12}\textminus{}0.02 & \cellcolor{gray!12}+0.23 & \cellcolor{gray!12}\textminus{}0.25 & \cellcolor{gray!12}\textbf{84.3 $\to$ 88.6} & \cellcolor{gray!12}\textminus{}0.3 & \cellcolor{gray!12}\textbf{+6.5} \\
 & \cellcolor{gray!12}\textcolor{gray}{XLS-R-SLS$^\dagger$} & \cellcolor{gray!12}\textminus{}0.04 & \cellcolor{gray!12}0.00 & \cellcolor{gray!12}\textminus{}0.05 & \cellcolor{gray!12}50.8 $\to$ 51.0 & \cellcolor{gray!12}+0.1 & \cellcolor{gray!12}+0.4 \\
 & \cellcolor{gray!12}\textcolor{gray}{XLS-R-Mamba$^\dagger$} & \cellcolor{gray!12}\textminus{}0.02 & \cellcolor{gray!12}+0.11 & \cellcolor{gray!12}\textminus{}0.13 & \cellcolor{gray!12}56.2 $\to$ 56.7 & \cellcolor{gray!12}+0.2 & \cellcolor{gray!12}+0.9 \\
 & \cellcolor{gray!12}\textcolor{gray}{ProsDD$^\dagger$} & \cellcolor{gray!12}+0.14 & \cellcolor{gray!12}\textbf{+0.83} & \cellcolor{gray!12}\textbf{\textminus{}0.68} & \cellcolor{gray!12}\textbf{50.0 $\to$ 56.6} & \cellcolor{gray!12}\textminus{}3.3 & \cellcolor{gray!12}\textbf{+10.2} \\
 & \cellcolor{gray!12}\textcolor{gray}{SLIM$^\dagger$} & \cellcolor{gray!12}0.00 & \cellcolor{gray!12}\textbf{+0.39} & \cellcolor{gray!12}\textbf{\textminus{}0.40} & \cellcolor{gray!12}\textbf{74.3 $\to$ 77.8} & \cellcolor{gray!12}\textminus{}1.5 & \cellcolor{gray!12}\textbf{+10.5} \\
 & \cellcolor{gray!12}\textcolor{gray}{Alethia$^\dagger$} & \cellcolor{gray!12}+0.06 & \cellcolor{gray!12}\textbf{+0.92} & \cellcolor{gray!12}\textbf{\textminus{}0.86} & \cellcolor{gray!12}\textbf{56.4 $\to$ 62.6} & \cellcolor{gray!12}\textminus{}0.9 & \cellcolor{gray!12}\textbf{+9.8} \\
\bottomrule
\end{tabular}
\caption{Full results. $\Delta$: mean score of \AB{} minus \AA{} within a class (negative: transitioning speech looks more bona fide). $\Delta_{\text{SP}}{-}\Delta_{\text{BF}}$
measures the change in signed mean score
separation between synthetic and bona fide speech. EER on \AA{} and on \AB{}. $\Delta$FNR / $\Delta$FPR: change in spoof miss rate and bona fide false-alarm rate from \AA{} to \AB{}, with the threshold fixed at the EER point on \AA{}. Bold: 95\% cluster-bootstrap CI excludes zero (for EER, of the change). Grey, $^\dagger$: detector inverted on that corpus (pooled EER $>50\%$, Table~\ref{tab:auc}).}
\label{tab:full}
\end{table*}

\section{Naturalistic Controls}
\label{sec:appendix-nat}

We conduct additional analyses on bona fide
speech from IEMOCAP and MSP-Podcast to
examine whether the observed score shifts
persist after accounting for differences
in speech characteristics and emotional
trajectories.
Table~\ref{tab:nat} reports both adjusted
score shifts and within-utterance
arousal associations.

\paragraph{Adjusted score shifts.}
We estimate the association between
emotion condition and detector scores
using ordinary least squares (OLS)
regression.
The dependent variable is the
utterance-level logit spoof score,
and the main predictor distinguishes
\AB{} from \AA{}.
We include speaker fixed effects,
log utterance duration, and VAD annotations
from both speech regions as covariates.
The coefficient of the \AB{} indicator
provides the adjusted score shift.
We report speaker-clustered 95\%
confidence intervals.

These adjustments account for measured
differences in speaker, duration, and
regional emotional states while preserving
the comparison between emotion-stable
and emotion-changing bona fide speech.

\paragraph{Within-utterance arousal associations.}
We additionally examine how score changes
between the two speech regions relate to
changes in arousal.
For each utterance, we regress window-level
spoof scores on the proportion of each window
overlapping the second region.
The resulting coefficient estimates the
score difference between the second and
first regions.

We then regress this within-utterance
score difference on the signed change
in normalized arousal between regions,
pooling \AA{} and \AB{} utterances.
Slopes are reported per 0.1 increase
in normalized arousal, with confidence
intervals estimated using 1,000
speaker-level bootstrap resamples.

\begin{table}[!htbp]
\centering
\small
\setlength{\tabcolsep}{3pt}
\resizebox{\columnwidth}{!}{%
\begin{tabular}{@{}l rr rr rr@{}}
\toprule
 & \multicolumn{2}{c}{$\Delta_{\text{BF}}$, IEMOCAP} & \multicolumn{2}{c}{$\Delta_{\text{BF}}$, MSP} & \multicolumn{2}{c}{Arousal slope} \\
\cmidrule(lr){2-3}\cmidrule(lr){4-5}\cmidrule(l){6-7}
Detector & raw & adj. & raw & adj. & IEM. & MSP \\
\midrule
RawNet2 & +1.54 & \textbf{+1.35} & +0.13 & \textminus{}0.22 & \textbf{+0.55} & \textbf{+0.78} \\
AASIST & +1.66 & \textbf{+1.38} & +0.23 & \textminus{}0.05 & \textbf{+0.98} & \textbf{+0.70} \\
XLS-R-SLS & \textminus{}0.60 & \textbf{\textminus{}0.60} & +0.00 & \textbf{\textminus{}0.31} & 0.00 & +0.14 \\
XLS-R-Mamba & \textminus{}0.75 & \textbf{\textminus{}0.82} & +0.11 & \textbf{\textminus{}0.22} & +0.07 & \textbf{+0.12} \\
ProsDD & \textminus{}0.27 & 0.00 & +0.83 & \textbf{+0.54} & \textbf{\textminus{}0.50} & \textbf{+0.19} \\
SLIM & +0.14 & \textbf{+0.26} & +0.39 & \textbf{+0.24} & \textbf{\textminus{}0.29} & \textbf{+0.10} \\
Alethia & +0.79 & \textbf{+1.01} & +0.92 & \textbf{+0.59} & \textbf{\textminus{}0.75} & \textbf{+0.45} \\
\bottomrule
\end{tabular}}
\caption{Naturalistic bona fide controls. $\Delta_{\text{BF}}$ raw and adjusted for speaker, duration, and regional VAD; arousal slope: within-utterance second-minus-first-region score per 0.1 rise in normalized arousal. Bold: 95\% CI excludes zero.}
\label{tab:nat}
\end{table}

RawNet2 and AASIST are the only detectors
with significantly positive arousal slopes
on both IEMOCAP and MSP-Podcast.
In contrast, ProsDD, SLIM, and Alethia
show significant associations in opposite
directions across the two datasets.
These findings indicate that sensitivity
to arousal changes varies across detectors
and recording conditions.

\section{Detector Settings}
\label{sec:appendix-setup}

Table~\ref{tab:detectors} summarizes the
checkpoint sources and window configurations
used for inference.
All detectors are evaluated using
checkpoints trained or fine-tuned for
spoof detection on ASVspoof 2019 LA.
Audio is converted to 16\,kHz mono,
and overlapping windows are aggregated
to obtain utterance-level spoof scores.

\begin{table}[t]
\centering
\small
\resizebox{\columnwidth}{!}{%
\begin{tabular}{@{}llrr@{}}
\toprule
Detector & Checkpoint & Window & Hop \\
\midrule
RawNet2 & official & 64{,}600 (4.04\,s) & 32{,}300 \\
AASIST & official & 64{,}600 (4.04\,s) & 32{,}300 \\
XLS-R-SLS & official & 64{,}600 (4.04\,s) & 32{,}300 \\
XLS-R-Mamba & official & 66{,}800 (4.18\,s) & 33{,}400 \\
ProsDD & official & 64{,}000 (4.00\,s) & 32{,}000 \\
SLIM & fine-tuned & 48{,}000 (3.00\,s) & 24{,}000 \\
Alethia & fine-tuned & 48{,}000 (3.00\,s) & 24{,}000 \\
\bottomrule
\end{tabular}}
\caption{
Detector inference configurations.
Window and hop sizes are reported in
samples at 16\,kHz.
The first five detectors use released
checkpoints, while SLIM and Alethia use
ASVspoof 2019 LA fine-tuned checkpoints.
}
\label{tab:detectors}
\end{table}

\section{Additional Analysis of Splicing Artifacts}
\label{sec:appendix-splice}

\begin{figure*}[t]
\centering
\includegraphics[width=\textwidth]
{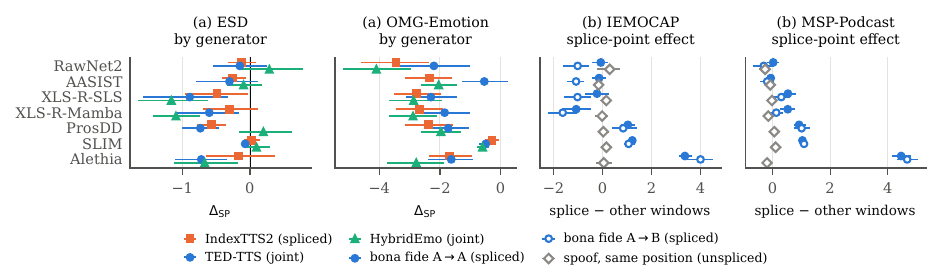}
\caption{
Analysis of splicing artifacts.
(a) Synthetic spoof score shifts from
emotion-stable (\AA{}) to emotion-changing
(\AB{}) speech, grouped by generator
on ESD and OMG-Emotion.
(b) Difference between spoof scores
from windows containing a splice boundary
and other windows within the same
utterance on IEMOCAP and MSP-Podcast.
Unspliced EmoTra-TTS speech is included
as a reference at the corresponding
position.
Positive values indicate higher spoof
scores near splice boundaries.
Error bars show 95\% bootstrap
confidence intervals.
}
\label{fig:splice}
\end{figure*}

We further investigate whether acoustic
discontinuities introduced by concatenation
could account for the observed differences
between emotion-stable and emotion-changing
speech (Figure~\ref{fig:splice}).

\paragraph{Generator-specific score shifts.}
Figure~\ref{fig:splice}a compares synthetic
score shifts across IndexTTS2, TED-TTS,
and HybridEmo on ESD and OMG-Emotion.
IndexTTS2 generates two speech segments
separately and concatenates them, whereas
TED-TTS and HybridEmo generate complete
utterances without splicing.
Significant negative score shifts also
appear in the unspliced generations,
indicating that concatenation is not
necessary for the observed reductions
in spoof scores.

\paragraph{Local splice-boundary effects.}
Figure~\ref{fig:splice}b compares spoof
scores from windows containing the
splice boundary with scores from other
windows in the same utterance.
We perform this analysis on IEMOCAP
and MSP-Podcast bona fide speech,
using unspliced EmoTra-TTS utterances
as a reference at the corresponding
temporal position.

Some detectors assign higher spoof scores
to windows containing splice boundaries.
However, the average boundary effects
are similar between \AA{} and \AB{}
bona fide speech.
These observations indicate that
splicing can influence local scores,
but its presence alone does not fully
account for the condition-level shifts.

\end{document}